\documentclass[12pt,preprint]{aastex}
\usepackage{natbib}
\usepackage{amssymb,amsmath}
\usepackage[colorlinks,
            linkcolor=blue,
            anchorcolor=blue,
            citecolor=blue,
            ]{hyperref}
\usepackage{lineno}

\begin{document}

\title{Radioactively-Powered Gamma-Ray Transient Associated with a Kilonova from Neutron Star Merger}
\author{Meng-Hua Chen$^{1}$, Rui-Chong Hu$^{1}$, En-Wei Liang$^{*1}$}
\altaffiltext{1}{Guangxi Key Laboratory for Relativistic Astrophysics, School of Physical Science and Technology, Guangxi University, Nanning 530004, People's Republic of China; lew@gxu.edu.cn}

\begin{abstract}
Association of GW170817/GRB170817A/AT2017gfo provides the first direct evidence for neutron star mergers as significant sources of $r$-process nucleosynthesis. A gamma-ray transient (GRT) would be powered by the radioactive decay of the freshly-synthesized $r$-process elements. By analyzing the composition and gamma-ray opacity of the kilonova ejecta in details, we calculate the lightcurve and spectrum of the GRT for a range of spherically symmetric merger ejecta models with mass $M_{\rm ej}=0.001 \sim 0.05M_{\odot}$ and expansion velocity $v_{\rm ej}= 0.1\sim 0.4c$. It is found that the peak of the GRT lightcurve depends on $M_{\rm ej}$ and $v_{\rm ej}$ as $t_{\rm pk} \approx 0.5~{\rm days} ~ (M_{\rm ej}/0.01M_{\odot})^{1/2}(v_{\rm ej}/0.1c)^{-1}$ and $L_{\rm pk} \approx 2.0\times10^{41} ~{\rm erg~s} ^{-1} (M_{\rm ej}/0.01M_{\odot})^{1/2}(v_{\rm ej}/0.1c)$. Most radiating photons are in the $100-3000$ keV band and the spectrum peaks at about 800~keV for different nuclear physics inputs. The line features are blurred out by the Doppler broadening effect. Adopting the ejecta parameters reported in literature, we examine the detection probability of the possible GRT associated with AT2017gfo. We show that the GRT cannot be convincingly detected neither with current nor with the proposed missions in the MeV band, such as ETCC and AMEGO. The low gamma-ray flux, together with the extremely low event rate at local universe, makes a discovery of GRTs a great challenge.
\end{abstract}


\section{Introduction}
\label{intro}

Kilonova powered by neutron-rich ejecta from merger of binary neutron stars was theoretically predicted and observationally confirmed \citep{Lattimer1974,Abbott2017a,Abbott2017b}.
Roughly half of the elements heavier than iron are synthesized through the rapid neutron capture process ($r$-process; \citealp{Burbidge1957}).
For the heaviest of them, neutron star mergers have been proposed as a promising site of heavy r-process nucleosynthesis \citep{Lattimer1974}.
With the recent multimessenger observation of AT2017gfo \citep{Abbott2017b}, neutron star mergers can be considered as the \emph{primary sites} of heavy r-process, at least in the current Universe.
The radioactive decay of $r$-process elements produces a characteristic electromagnetic transient in the ultraviolet/optical/near-infrared wavelengths, which is known as a ``kilonova'' (\citealp{Li1998,Metzger2010,Korobkin2012,Barnes2013,Kasen2013,Barnes2016}; see \citealp{Metzger2019} for a recent review).
The first unambiguous kilonova, named AT2017gfo, was discovered on 2017 August 17 \citep{Abbott2017b,Arcavi2017,Chornock2017,Coulter2017,Cowperthwaite2017,Drout2017,Evans2017,Kasliwal2017,Nicholl2017,Pian2017,Smartt2017,Tanvir2017}.
This kilonova followed a gravitational-wave signal compatible with the inspiral and coalescence of binary neutron stars (GW170817, \citealp{Abbott2017a}).
The luminosity and color evolution of AT2017gfo suggest the presence of two distinct components: an early ``blue'' component driven by wind ejecta and a later ``red'' component driven by the dynamical ejecta \citep{Kasen2017,Kilpatrick2017,Nicholl2017,Tanaka2017,Tanvir2017,Troja2017,Villar2017}.
The blue component is attributed to the ejecta dominated by light $r$-process nuclei (atomic mass number $A\lesssim140$), while the red component is likely from the ejecta rich in lanthanides and heavy $r$-process material ($A\gtrsim140$).
These observational features are well consistent with the theoretical expectation.
The absorption features in the kilonova spectrum may be used to identify specific heavy elements.
However, due to the large Doppler broadening and blend of many absorption lines, no individual heavy elements have been identified except for strontium (atomic number $Z=38$) in the observed spectra of AT2017gfo \citep{Watson2019}.

Besides the kilonova, a gamma-ray transient (GRT) would be also powered by the radioactive decay of freshly-synthesized heavy elements \citep{Hotokezaka2016}.
Radioactive decay of unstable nuclei are usually left the daughter nucleus in an excited state.
The subsequent decay of the daughter nucleus to a lower-energy state results in gamma-ray emission in specific energy.
Detection of these gamma-ray photons can provide conclusive evidence for identifying individual elements and tracking their evolution.
Radioactive gamma-ray emission from neutron star mergers was studied by several groups.
They focused on identifying the gamma-ray line features associated with $r$-process elements \citep{Li2019,Wu2019a,Korobkin2020,Chen2021}.
For instance, \cite{Chen2021} showed that the decay chain of $^{132}$Te ($t_{1/2}=3.21$~days)$\rightarrow$$^{132}$I ($t_{1/2}=0.10$~days)$\rightarrow$$^{132}$Xe would produce several bright gamma-ray lines in the MeV band and would be marginally detectable with the future gamma-ray detector if the source is at a distance of 40~Mpc.
Note that the characteristics of a GRT and associated kilonova are sensitive to the ejecta properties, especially the mass and expansion velocity.
Recent numerical simulations of binary neutron star mergers show that the ejecta mass may be in a range from $\sim0.001M_{\odot}$ to nearly $0.05M_{\odot}$ (the ejecta mass of neutron star black hole merger can be even higher, up to $\sim0.3M_{\odot}$ \citealp{Rosswog2005}), which depends strongly on the total mass, mass ratio of binary system, and neutron star equation of state \citep{Bauswein2013,Hotokezaka2013,Sekiguchi2016,Dietrich2017,Fujibayashi2018,Radice2018,Shibata2019,Nedora2021}. The typical velocity of the ejecta is in the sub-relativistic regime, $\sim0.1-0.4c$, where $c$ is the speed of light. Investigation of the GRTs and kilonovae in such plausible ranges would be helpful for assessing their delectability with current and upcoming gamma-ray missions.

The gamma-ray opacities of $r$-process elements are needed to derive the lightcurve and spectrum of a GRT. \cite{Korobkin2020} studied the gamma-ray emission by simply assuming that all heavy $r$-process elements have an opacity comparable to the iron element. Since the probability of photoelectric absorption is proportional to the atomic number as $Z^{4}$, the opacities of $r$-process elements are significantly higher than those of the iron element in the range where photoelectric absorption dominates. The higher opacities, the more gamma-ray photons will suffer absorption, hence dimmer resulting luminosity.

This paper studies the GRTs from neutron star mergers in reasonable ranges based on numerical results by considering the opacities of $r$-process elements in detail, in which the opacity is estimated based on the detailed nuclide composition obtained from nucleosynthesis calculations of the $r$-process. In Section~\ref{sec2}, we study the composition and gamma-ray opacity in detail. Section~\ref{sec3} presents the model and our numerical results for deriving the GRTs for a grid of models. As a case study, we apply our model to the possible GRT associated with AT2017gfo for evaluating its detection probability in Section~\ref{sec4}. Conclusions and discussion are presented in Section~\ref{sec5}.

\section{Composition and Gamma-ray Opacity of the Merger Ejecta}
\label{sec2}

\subsection{Ejecta Model}
\label{sub1}

We assume the merger ejecta is spherical and expanding homogenously, and the ejected material is a sum of two components, i.e. a neutron-rich dynamical ejecta generated by tidal disruption and a wind ejecta driven by the post-merger remnant. The ejecta model is described as follows.
The merger ejecta is divided into $N$ ($N=100$) layers in our calculation.
The expansion velocity of the $n$th layer is given as $v_{n}=v_{0}+n\Delta v$, where $n=1,2,...,99$, $v_0$ is the velocity of the $0$th layer, and $\Delta v=0.001c$.
The radius of the $n$th layer at time $t$ is $R_n=v_n t$. Its mass is
\begin{equation}
 M_n=\int_{R_{n-1}}^{R_{n}}4\pi \rho r^2dr,
 \end{equation}
where the mass density profile is parameterized as a power-law \citep{Zhu2021},
\begin{equation}
\rho(v_n, t)=\rho_0(t)\left(\frac{v_n}{v_0}\right)^{-3}.
\end{equation}
For a given mass of the ejecta ($M_{\rm ej}$), $\rho_0(t)$ is derived from
\begin{equation}
M_{\rm ej} = \int_{R_{\rm 0}}^{R_{\rm N-1}}4\pi\rho(v_n,t)r^2dr.
\end{equation}
The characteristic ejecta velocity ($v_{\rm ej}$) can be estimated as
\begin{equation}
v_{\rm ej}=(2E_{\rm kin}/M_{\rm ej})^{1/2}
\end{equation}
where  $E_{\rm kin}\equiv \sum_n M_n v_n^2/2$ is the sum of the kinetic energy of all layers.

In our calculation, we vary $M_{\rm ej}$ from $0.001M_{\odot}$ to $0.05M_{\odot}$ with a bin size of $0.001M_{\odot}$ and vary $v_0$ from $0.05c$ to $0.35c$ with a bin size of $0.01c$. The derived $v_{\rm ej}$ value ranges in $0.1-0.4c$.
The $M_{\rm ej}$ and $v_{\rm ej}$ cover the ranges of ejecta parameters reported in the literature \citep{Bauswein2013,Hotokezaka2013,Sekiguchi2016,Dietrich2017,Fujibayashi2018,Radice2018,Shibata2019,Nedora2021}.

\subsection{Composition}
\label{sub2}

To obtain the detailed composition of $r$-process materials, we employ the nuclear reaction network code SkyNet for $r$-process simulations\citep{Lippuner2015,Lippuner2017}.
The network includes 7843 nuclides and $\sim140,000$ nuclear reactions, and use nuclear reaction rates from the JINA REACLIB database \citep{Cyburt2010}.
In our calculations, we use the state-of-the-art nuclear physics inputs, including the experimental masses from the latest Atomic Mass Evaluation (AME2020, \citealp{Wang2021}) and experimental data from newly published database NUBASE2020 \citep{Kondev2021}.
For the nuclide species without experimental data, we use the theoretical data from Finite-Range Droplet Model (FRDM, \citealp{Moller2016}).
The neutron capture rates are calculated with each associated mass data using the nuclear reaction code TALYS \citep{Goriely2008}.
For spontaneous fission, neutron-induced fission and $\beta$-delayed fission, we use the double Gaussian fission fragment distributions from Kodama-Takahashi model (KT, \citealp{Kodama1975}).
We note that the choice of nuclear physics models and fission fragment distributions has significant impact on the $r$-process nucleosynthesis simulations \citep{Mumpower2018,Zhu2021}.

In our calculation, the composition of the dynamical ejecta is calculated with entropy $s=10~k_B$/baryon, expansion timescale $\tau_{\rm dyn}=10$~ms, and electron fraction $Y_{\rm e}\sim0.05-0.25$ (with increments of 0.01).
For a wind ejecta, we take $s=20~k_B$/baryon, $\tau_{\rm dyn}=30$~ms, and $Y_{\rm e}\sim0.20-0.40$ (with increments of 0.01).
Following \cite{Wu2019b}, we assume that the ejected material contains a Gaussian $Y_{\rm e}$ distribution characterized by a central value $Y_{\rm e, c}$ and a width $\Delta Y_{\rm e}$, i.e.,
\begin{equation}
Y_i(t)=\int Y_i(Y_{\rm e}) G(Y_{\rm e}\mid Y_{\rm e, c}, \Delta Y_{\rm e}^2)d Y_{\rm e},
\end{equation}
where $Y_i(Y_{\rm e})$ is the abundance of the $i$th nuclide in a given $Y_{\rm e}$ and $G(Y_{\rm e}\mid Y_{\rm e, c}, \Delta Y_{\rm e}^2)$ is the normalized Gaussian distribution.
We take $Y_{\rm e,c}=0.15$ for dynamical ejecta and $Y_{\rm e,c}=0.30$ for wind ejecta.
The width of the Gaussian distribution is set to be $\Delta Y_{\rm e}=0.04$ \citep{Wu2019b}.
The compositions of both dynamical and wind ejecta are kept fixed when we vary the ejecta mass and expansion velocity of each ejecta component.

\begin{figure}
\centering
\includegraphics[width=0.8\textwidth]{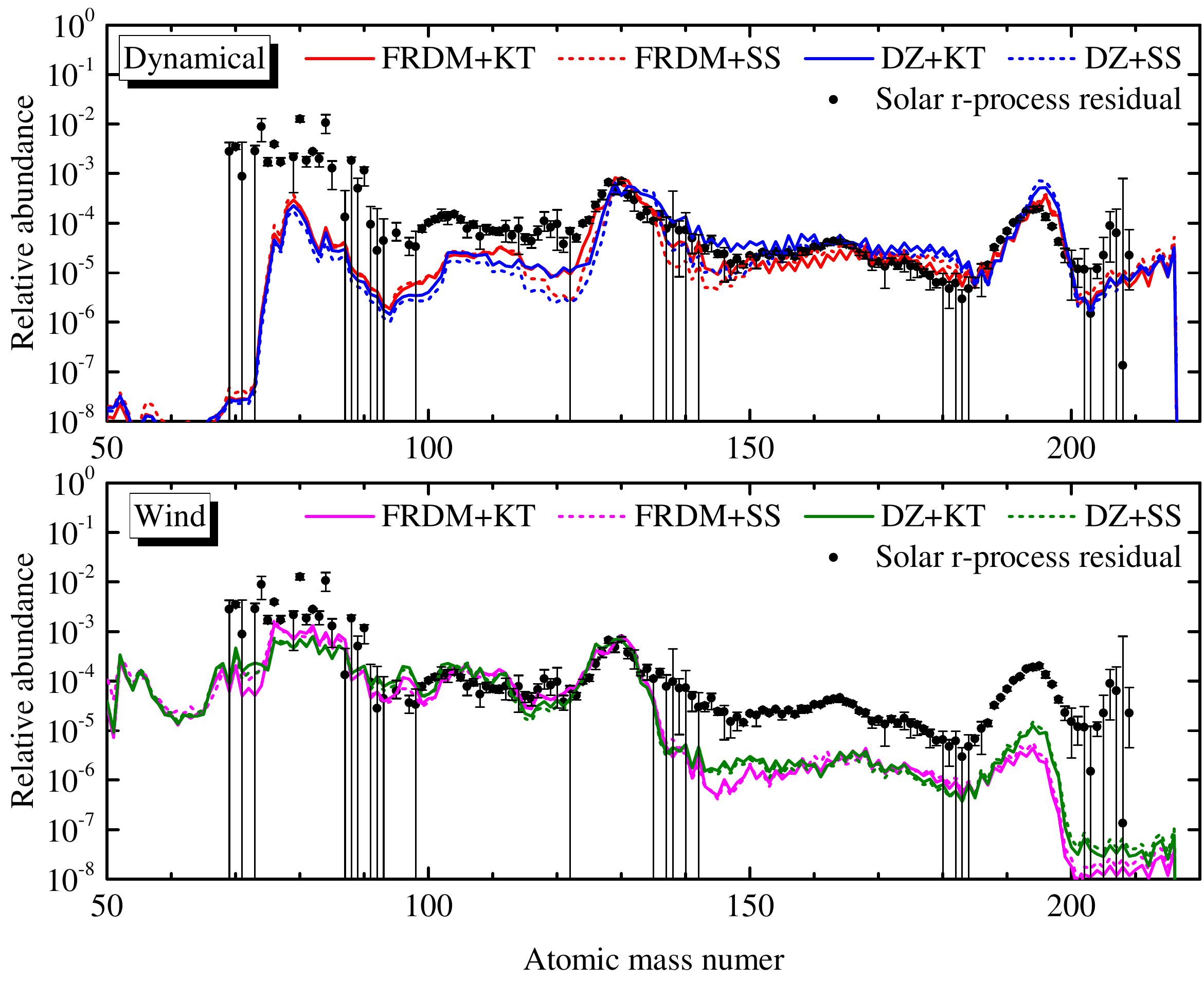}
\caption{Abundance patterns of $r$-process element for dynamical and wind ejecta at $t=10^9$~s. The solar $r$-process abundance taken from  \citep{Arnould2007} is shown with black dots for comparison.
FRDM = Finite-Range Droplet Model; DZ = Duflo-Zuker model; KT = Kodama-Takahashi fission model; SS = symmetric split model.}
\label{abundance}
\end{figure}

Figure~\ref{abundance} shows the resulting abundance patterns from $r$-process nucleosynthesis simulations.
It is found that the dynamical ejecta is sufficiently neutron rich to produce a wide range of elements from the second to third $r$-process peak.
For a high-$Y_{\rm e}$ wind ejecta, the $r$-process is not fully proceeded and only the nuclides around the first and second $r$-process peaks are produced.
It is known that the elemental abundances in some $r$-process enriched halo stars show remarkable agreement with the Solar $r$-process residuals, where the contribution of the $s$-process is subtracted \citep{Sneden2008,Roederer2017,Thielemann2017,Beniamini2018}.
This fact suggests that a single phenomenon reproduceds the solar-like r-process abundance patterns.
Hence we take the Solar $r$-process residuals from \cite{Arnould2007} for a comparison.
One can observe that the neutron star merger reproduces the main features of the Solar $r$-process residuals.
If the nucleosynthesis yields of AT2017gfo reproduce the Solar $r$-process residuals and can be taken as typical for neutron star mergers, together with the merger rate of $R_{\rm BNS}=1540^{+3200}_{-1220}$~Gpc$^{-3}$~yr$^{-1}$ inferred from the LIGO/Virgo discovery \citep{Abbott2017a}, one obtains that neutron star mergers are major sources of $r$-process elements in our Galaxy and Solar System \citep{Kasen2017,Cote2018,Hotokezaka2018}.

To explore the effects of nuclear physics inputs in the $r$-process nucleosynthesis on the resulting GRTs, we estimate the abundance patterns by using the nuclear masses from the Duflo-Zuker (DZ) model (\citealp{Duflo1995}) and the fission reactions from the symmetric split (SS) model (\citealp{Lippuner2015}). The results are shown in Figure~\ref{abundance}.
For a neutron-rich dynamical ejecta, the $r$-process proceeds past the third $r$-process peak and the fission properties play a key role in shaping the abundance patterns. Thus, the abundance patterns calculated with the double Gaussian fission model have a broader second $r$-process peak since its fission products are in a wider range.
For a high-$Y_{\rm e}$ wind ejecta, the abundance patterns is similar for different nuclear physics inputs, which may be an artifact of fixing the expansion timescale $\tau_{\rm dyn}=30$~ms.
This is reasonable since the $r$-process path {\bf is} closer to the valley of stability, where the discrepancies between the masses predicted by our selected nuclear mass models are modest \citep{Zhu2021}.

\subsection{Gamma-Ray Photon Opacity}
\label{sub3}

To calculate the radioactively-powered GRTs from neutron star mergers, we need to consider the effect of absorption and scattering by the ejected material.
The gamma-ray photons released by radioactive decay of unstable nuclei are in the energy range from a few keV to $\sim3$ MeV.
These MeV gamma-ray photons may interact with matter by three main mechanisms: photoelectric absorption, Compton scattering, and pair production.
In the photoelectric absorption process, the gamma-ray photon loses all of its energy in one interaction.
The probability for this process sensitively depends on gamma-ray energy $E_{\gamma}$ and atomic number $Z$.
In Compton scattering, the gamma-ray photon loses only part of its energy in one interaction.
The probability for this process is weakly dependent on $E_{\gamma}$ and $Z$.
The gamma-ray photon can lose all of its energy in one pair production interaction.
The three interaction processes described above all contribute to the opacity of merger ejecta.

\begin{figure}
\centering
\includegraphics[width=1.0\textwidth]{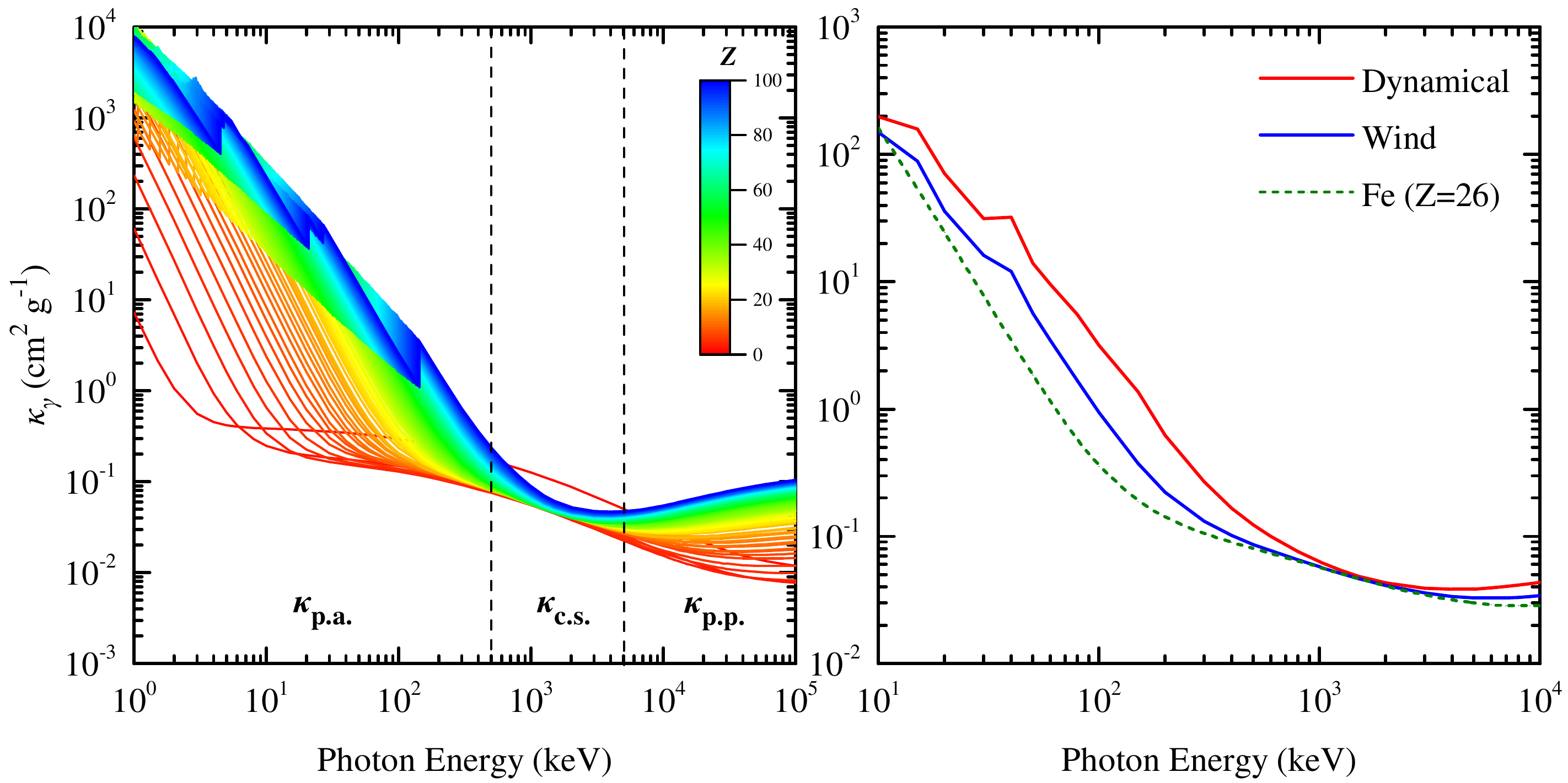}
\caption{Gamma-ray photon opacity as a function of photon energy for individual elements from hydrogen ($Z=1$) to fermium ($Z=100$) (left panel) and for the wind and dynamical ejecta (right panel). The gamma-ray opacity of the iron element is also shown with a dashed line in the right panel for comparison.
The opacity is calculated by considering the photoelectric absorption ($\kappa_{\rm p.a.}$), Compton scattering ($\kappa_{\rm c.s.}$), and pair production ($\kappa_{\rm p.p.}$) for the gamma-rays. The opacity values of these processes are taken from XCOM database.}
\label{opacity}
\end{figure}

We extract the opacity values of all elements from the XCOM database published by the National Institute of Standards and Technology (NIST) website\footnote{https://www.nist.gov/pml/xcom-photon-cross-sections-database}.
The opacity curves of the elements from hydrogen ($Z=1$) to fermium ($Z=100$) at energies between 1~keV and 100~MeV are shown in Figure~\ref{opacity}.
As the photon energy increases, the dominant interaction mechanism shifts from photoelectric absorption to Compton scattering to pair production.
For gamma-ray energies below $500$~keV, photoelectric absorption is the dominant process.
The probability for this process increases significantly for high-$Z$ elements.
For gamma-ray photons at energies between $500$~keV and $5$~MeV, Compton scattering becomes the dominant interaction in matter.
The probability for this process is weakly dependent on $Z$ and the opacity curves for all elements are nearly identical.
Pair production can occur with gamma-ray energies exceeding $1.022$~MeV and becomes a significant process at energies over $5$~MeV.

We calculate the total opacity based on the mass-fraction weighted {\bf average} of the ejecta, i.e.,
\begin{equation}
\kappa_{\gamma}(E_{\gamma})=\sum_i A_i Y_i(t) \kappa_i(E_{\gamma}),
\end{equation}
where $A_i$ is the atomic mass number of the $i$th nuclide, $Y_i(t)$ is the corresponding abundance, and $\kappa_i(E_{\gamma})$ is the opacity of the $i$th nuclide.
In Figure~\ref{opacity}, we show the opacities of the dynamical and wind ejecta, where dashed line shows the opacity of the iron element.
It can be noticed that the opacities are higher than that of the iron element in the sub-MeV energy range by a factor of $\sim4$ in the dynamical ejecta and $\sim2$ in the wind ejecta.
In particular, as shown in Figure~\ref{opacity}, the opacity of the dynamical ejecta is higher than that of the iron element by roughly an order of magnitude for photons at energy around 100~keV.
This is because the probability of photoelectric absorption process is substantially enhanced by high-$Z$ elements.
For photon energy $\gtrsim1$~MeV, the opacities of the dynamical and wind ejecta are very similar and vary relatively slowly with photon energy.

\section{Radioactively-Powered Gamma-Ray Transients}
\label{sec3}

\subsection{GRT Model}

Our calculations of radioactively-powered GRTs are implemented based on a radiation transfer model reported in \cite{Chen2021}.
The gamma-ray energy generation rate is given by summing the energy generation of all decay mode for all nuclides, i.e.,
\begin{equation}
\dot{E_{\gamma}}=N_B\sum_i Y_i(t) \sum_j\frac{\epsilon_{ij}}{\tau_{ij}},
\end{equation}
where $N_B$ is the total number of baryons, $\epsilon_{ij}$ is the total energy of the gamma rays generated in the $j$th decay mode of the $i$th nuclide, and $\tau_{ij}$ is the mean lifetime.
The gamma-ray radiation data in each decay mode (including the $\alpha$-decay and $\beta$-decay) are taken from the NuDat2 database at the National Nuclear Data Center\footnote{http://www.nndc.bnl.gov/nudat2/}.

To obtain the emitted gamma rays from the ejecta, we solve the radiative transfer equation to get the emission intensity $I_E$, i.e.,
\begin{equation}
\frac{dI_E}{dl}=-\alpha(E_{\gamma}) I_E+j_E,
\end{equation}
where $\alpha(E_{\gamma})$ is the absorption coefficient, $j_E$ is the emission coefficient, and
$l$ is the photon path length.
For the gamma-ray photons travel from $l_0$ to $l_m$, we can get the intensity $I_E$ to be
\begin{equation}
I_E=I_E(l_0)e^{(-\tau(l_m)-\tau(l_0))}+\int_{l_0}^{l_m}j_E e^{(-\tau(l_m)-\tau(l))}dl,
\end{equation}
where the optical depth $\tau(l)$ is given by $\tau(l)=\int\alpha(E_{\gamma})dl$.
Here we use the analytic formulas for optical depth as described in \cite{Chen2021}.
Then the observed flux contributed by the $n$th layer can be obtained by
\begin{equation}
F_n=\int I_E\cos\theta d\Omega
\approx 2\pi \int_{0}^{\theta} I_E\theta d\theta,
\end{equation}
where $\theta$ is the angle between line of sight and the line between ejecta centre and observer.
The total observed photon flux produced by the radioactive decay of $r$-process elements can be obtained by summarizing the contributions of all layers,
\begin{equation}
F_{\gamma}=\sum_n F_n.
\end{equation}

\subsection{Numerical Results}

\begin{figure}
\centering
\includegraphics[width=1.0\textwidth]{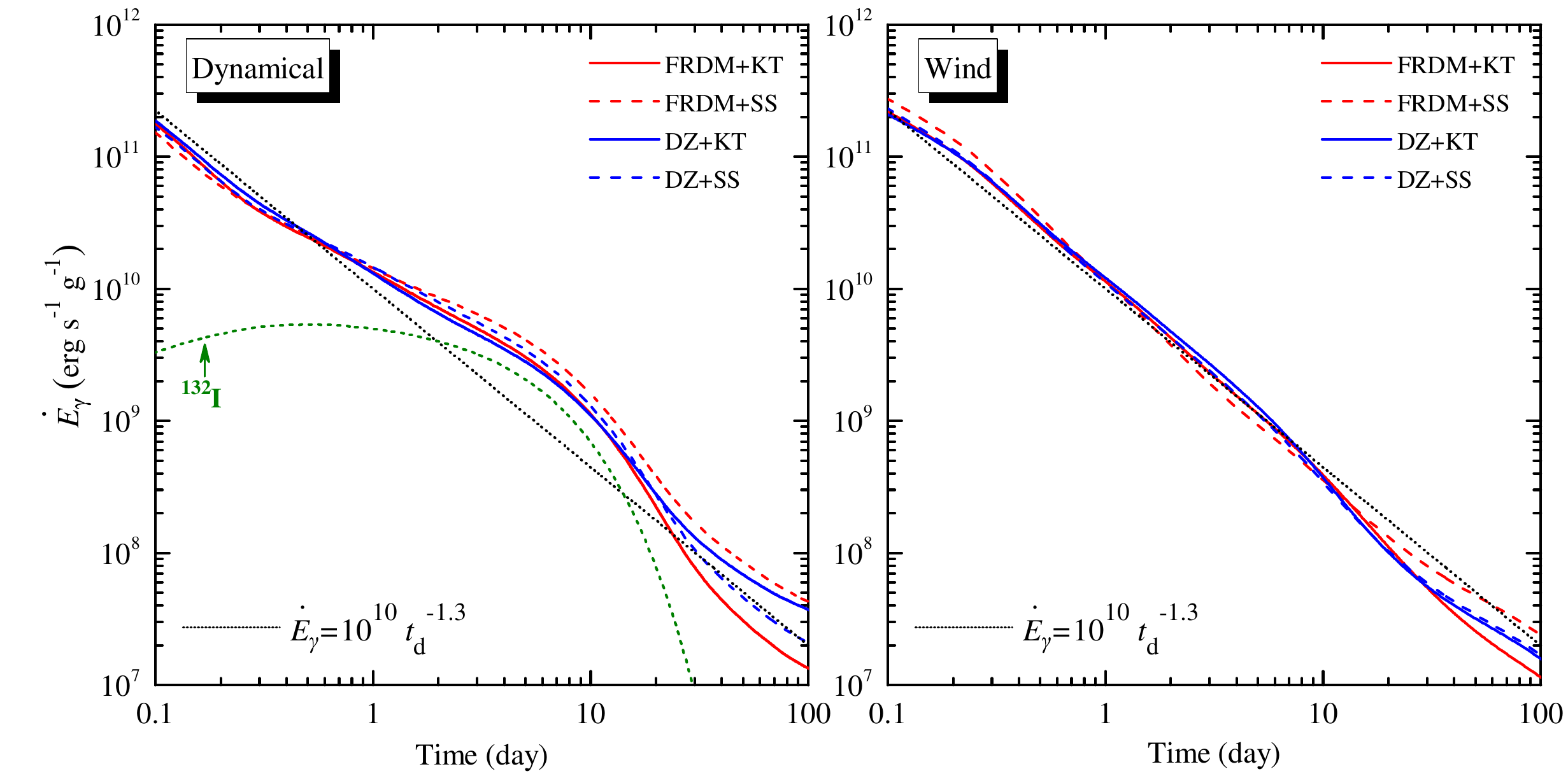}
\caption{Gamma-ray energy generation rates of the dynamical and wind ejecta.
The dotted lines indicate the power-law gamma-ray energy generation rate, i.e., $\dot{E}_{\gamma}=10^{10}t_{\rm d}^{-1.3}$~erg~s$^{-1}$~g$^{-1}$ ($t_{\rm d}$ is the time after the merger in days).
Green dashed line show the contribution from the $\beta$-decay of $^{132}$I.}
\label{energy}
\end{figure}

Figure~\ref{energy} shows the gamma-ray energy generation rates for the dynamical and wind ejecta, where the dotted line indicates the power-law energy generation rate, i.e., $\dot{E_{\gamma}}=10^{10}~t_{\rm d}^{-1.3}$~erg~s$^{-1}$~g$^{-1}$, where $t_{\rm d}$ is the time after the merger in days.
One can observe that the gamma-ray energy generation rate of the wind ejecta can be roughly described with a power-law function.
For the dynamical ejecta, a ``bump" feature is shown at several days after the merger. We calculate the generated gamma-ray energy of each nuclide in the dynamical ejecta and find that the bump feature is dominated by $^{132}$I since the radioactive decay of $^{132}$I releases $\sim2260$~keV gamma-ray energy per decay, being higher by a factor of $\sim3$ than the mean gamma-ray energy ($710$~keV) of 1309 nuclide species in NuDat2 database.

In Figure~\ref{energy}, we also show the gamma-ray energy generation rates calculated with different nuclear physics inputs.
For a high-$Y_{\rm e}$ wind ejecta, the gamma-ray energy generation rates are similar for different nuclear physics inputs due to their similar abundance pattern.
The gamma-ray energy generation rates in the dynamical ejecta are sensitive to the nuclear physics inputs.
This is because the $r$-process path for the very neutron-rich matter which constitutes the dynamical ejecta lies much farther away from the valley of stability, where nuclear physics uncertainties are higher.

\begin{figure}
\centering
\includegraphics[width=1.0\textwidth]{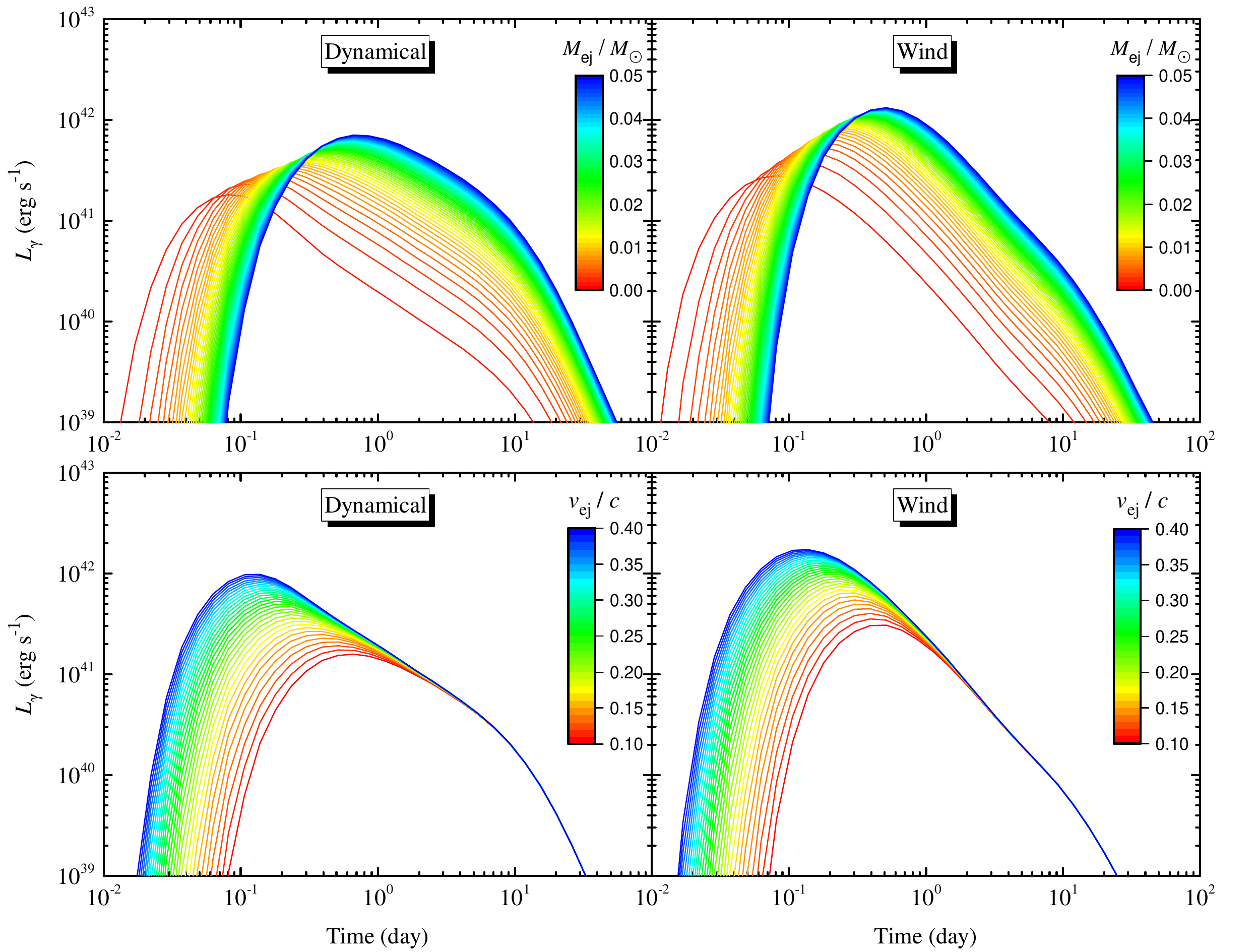}
\caption{Radioactively-powered GRT lightcurves of the dynamical and wind ejecta by varying the ejecta mass from $0.001$ to $0.05M_{\odot}$ (with increment of $0.001M_{\odot}$) and setting the ejecta expansion velocity at $v_{\rm ej}=0.2c$ (top two panels) and by varying the expansion velocities from $0.1$ to $0.4c$ (with increment of $0.01c$) and setting the ejecta mass at $M_{\rm ej}=0.01M_{\odot}$ (bottom two panels).}
\label{gamma}
\end{figure}

\begin{figure}
\centering
\includegraphics[width=1.0\textwidth]{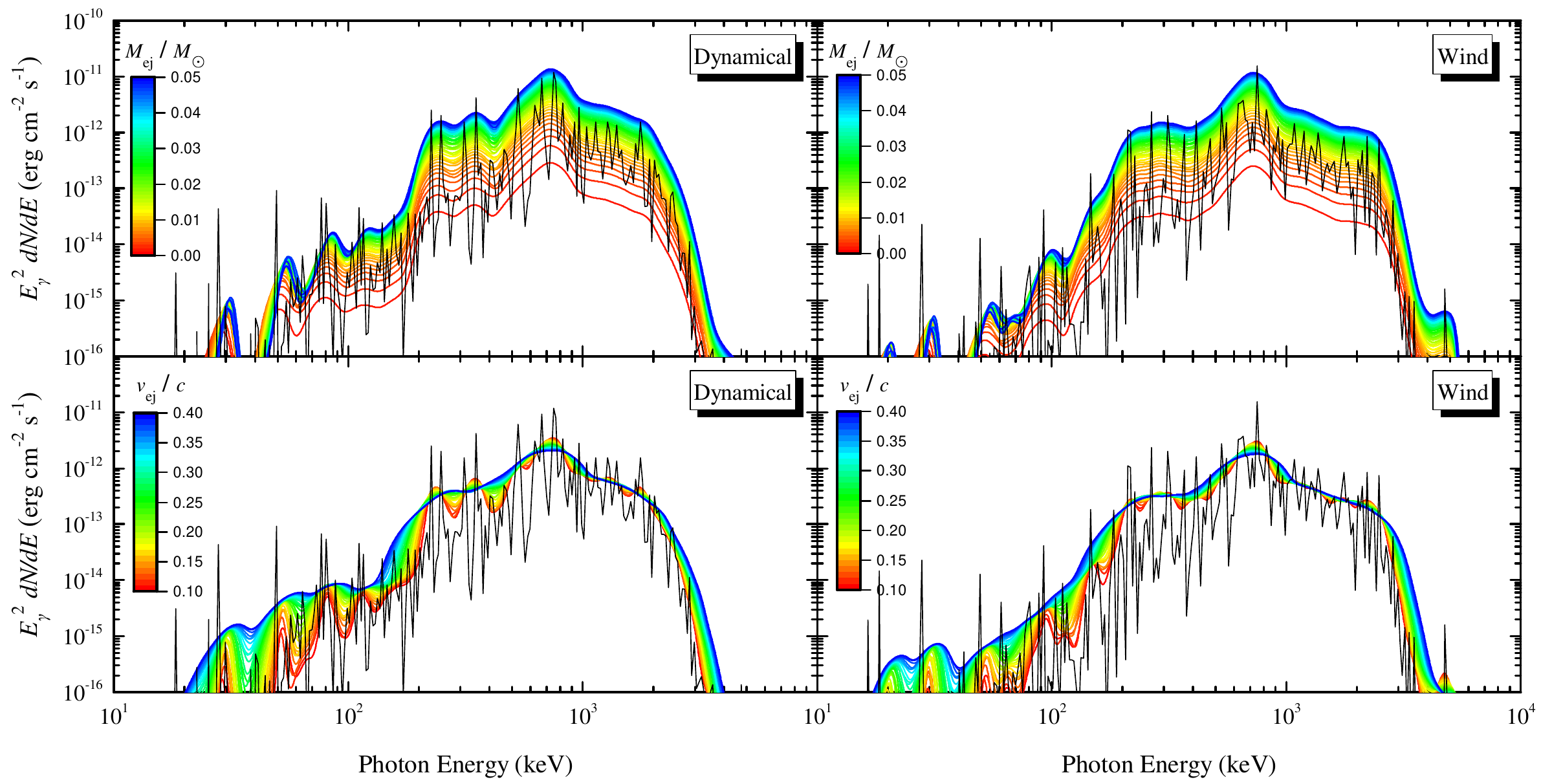}
\caption{The same as Figure~\ref{gamma}, but for the energy flux spectra at $t=1$~day.
Black lines indicate the spectrum produced by the radioactive decay of $r$-process elements at the rest frame and $M_{\rm ej}$ is set to $0.01M_{\odot}$.
Distance to the source is $40$~Mpc.}
\label{spectra}
\end{figure}

Figures~\ref{gamma} and \ref{spectra} show the derived GRT lightcurves and spectra of the dynamical and wind ejecta from our model by varying the ejecta mass from $0.001$ to $0.05M_{\odot}$ (with increment of $0.001M_{\odot}$) and velocity from $0.1$ to $0.4c$ (with increment of $0.01c$).
For the same ejecta mass and expansion velocity, the GRT lightcurves of the dynamical and wind ejecta are almost identical. This is because the gamma-ray energy generation rate only differs within a factor of $\sim2$.
The GRT lightcurves of both the dynamical and wind ejecta are sensitive to the mass and the velocity.
The GRT lightcurve of an ejecta with higher mass have a higher peak luminosity, peaks at later time, and lasts longer for a given $v_{\rm ej}$.
Their peak luminositiers vary from $10^{40}\sim 10^{42}$ erg s$^{-1}$.
The early GRT lightcurve ($t\lesssim3$ days) is sensitive to $v_{\rm ej}$, but it is degenerated at the late time for different $v_{\rm ej}$. This is reasonable since the ejecta at the peak time is almost transparent to gamma-ray photons, and the lightcurves for different velocities tend to be degenerated due to their same ejecta mass.
The GRT lightcurves have little sensitivity {\bf to} the opacity {\bf due} to the spectrum peak {\bf falling} on the energy range in which photoelectric absorption is subdominant (see Figure~\ref{spectra}).
The dependence of the GRT peak time and peak luminosity generally follows the relations
\begin{equation}
\label{tpk}
t_{\rm pk} \approx 0.5~ {\rm days} ~\left(\frac{M_{\rm ej}}{0.01M_{\odot}}\right)^{1/2}\left(\frac{v_{\rm ej}}{0.1c}\right)^{-1},
\end{equation}
\begin{equation}
\label{lpk}
L_{\rm pk} \approx 2.0\times10^{41}~~{\rm erg}~{\rm s}^{-1}~\left(\frac{M_{\rm ej}}{0.01M_{\odot}}\right)^{1/2}\left(\frac{v_{\rm ej}}{0.1c}\right).
\end{equation}

The energy flux spectra of the dynamical and wind ejecta at $t=1$~day are shown in Figure~\ref{spectra}.
The black lines in each panel indicate the gamma-ray lines at the rest frame and $M_{\rm ej}$ is set to $0.01M_{\odot}$.
The energies of emitted gamma-ray photons are typically in a range from $100-3000$~keV, and most of the gamma-ray photons are at energies of $\sim800$~keV.
The low-energy ($E_{\gamma}\lesssim200$~keV) gamma-ray photons seriously suffer the photoelectric absorption by high-$Z$ elements in the ejecta (as discussed in Section~\ref{sub3}) and only the gamma rays emitted from the near side of sphere can be seen.
From Figure~\ref{spectra} we see that the spectrum of an ejecta with higher mass have a larger energy flux.
The peak fluxes at $\sim800$~keV vary from $10^{-13}$ to $10^{-11}$~erg~cm$^{-2}$~s$^{-1}$.
A higher expansion velocity leads to a flatter spectrum due to the Doppler broadening effect.

\begin{figure}
\centering
\includegraphics[width=0.8\textwidth]{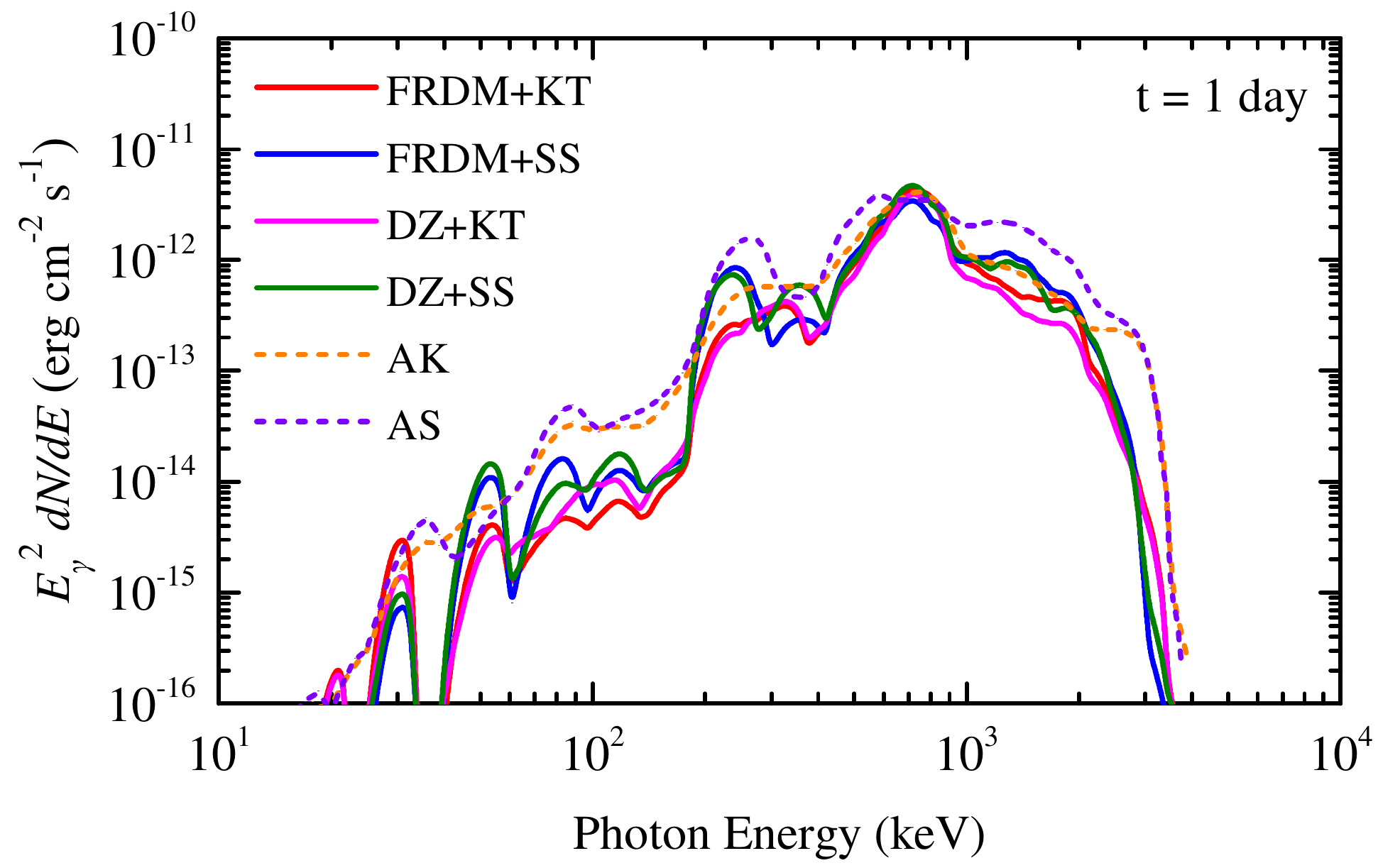}
\caption{The energy flux spectra calculated with different nuclear physics inputs as marked with different color solid lines in the figure, assuming the ejecta parameters as $M_{\rm ej}=0.0065M_{\odot}$, $v_{\rm ej}=0.2c$, and $Y_{\rm e}=0.05$, which are similar to the basic composition used in \cite{Korobkin2020}. The results for axisymmetric configuration models taken from \cite{Korobkin2020} are also presented with dashed lines, where ``AS" is for the axisymmetric plus symmetric split fission model and ``AK" is for the axisymmetric plus Kodama-Takahashi fission model. The distance is scaled to the host galaxy of GW170817, i.e., $D=40$~Mpc. }
\label{model}
\end{figure}

The nuclear physics inputs may affect the magnitude and shape of the gamma-ray spectrum.
To explore the sensitivity to the nuclear physics inputs, we calculate the energy flux spectra by using different nuclear mass models and fission fragment distributions. Our results are shown in Figure~\ref{model}, where properties of the merger ejecta are taken as $M_{\rm ej}=0.0065M_{\odot}$, $v_{\rm ej}=0.2c$, and $Y_{\rm e}=0.05$, which is similar to the basic compositions used in \cite{Korobkin2020}.
It is found that the spectra calculated by using different nuclear physics have very similar magnitude and shapes, although the peak fluxes have some subtle differences.
We note that our calculation assumed a spherical expanding ejecta model, which may cause an underestimate of the gamma-ray emission.
Recently, \cite{Korobkin2020} applied an axisymmetric model in the gamma-ray spectrum calculations (see their Figure~4).
In Figure~\ref{model}, we compare our resulting spectra to those calculated using axisymmetric model from \cite{Korobkin2020}.
The distance is rescaled to the host galaxy of GW170817, i.e., $D=40$~Mpc.
One can observe that the energy flux spectra calculated by using a spherical symmetric model and an axisymmetric model have similar global shapes.
The spectral peak at about 800~keV is also showed in the results of \cite{Korobkin2020}.
The non-spherical geometry of ejecta produces a slightly larger flux than that of the spherical geometry. This discrepancy may be caused by their use of the gamma-ray opacity of iron element, or Monte Carlo radiative transport code.

\section{Application to Kilonova AT2017gfo}
\label{sec4}

Kilonova AT2017gfo provides a solid case for studying its associated GRT. We apply our GRT model to this unique case.
To estimate the GRT of the merger event like AT2017gfo, one should know the ejecta parameters, i.e., ejecta mass and expansion velocity.
Several groups found that the lightcurve of kilonova AT2017gfo is well fit with the two component model, and derive $M_{\rm ej}\sim0.025M_{\odot}$  and $v_{\rm ej}\sim0.3c$ for the wind ejecta and $M_{\rm ej}\sim0.035M_{\odot}$ and $v_{\rm ej}\sim0.1c$ for the dynamical ejecta \citep{Arcavi2017,Chornock2017,Cowperthwaite2017,Kasen2017,Kilpatrick2017,Nicholl2017,Tanaka2017}.
Alternately, \cite{Tanvir2017} and \cite{Troja2017} fitted the lightcurve of AT2017gfo with an axisymmetric radiative transfer models from \cite{Wollaeger2018}, yielding   $M_{\rm ej}\sim0.015M_{\odot}$ and $v_{\rm ej}\sim0.08c$ for the wind ejecta and $M_{\rm ej}\sim0.002M_{\odot}$ and $v_{\rm ej}\sim0.2c$ for the dynamical ejecta.
We adopt these parameter sets as inputs to calculate the spectrum of the GRT associated with AT2017gfo.
When combining two ejecta components, we simply add the weighted compositions and ignore potential hydrodynamical interaction between the dynamical and wind ejecta in our calculations \citep{Kawaguchi2018}.

\begin{figure}
\centering
\includegraphics[width=1.0\textwidth]{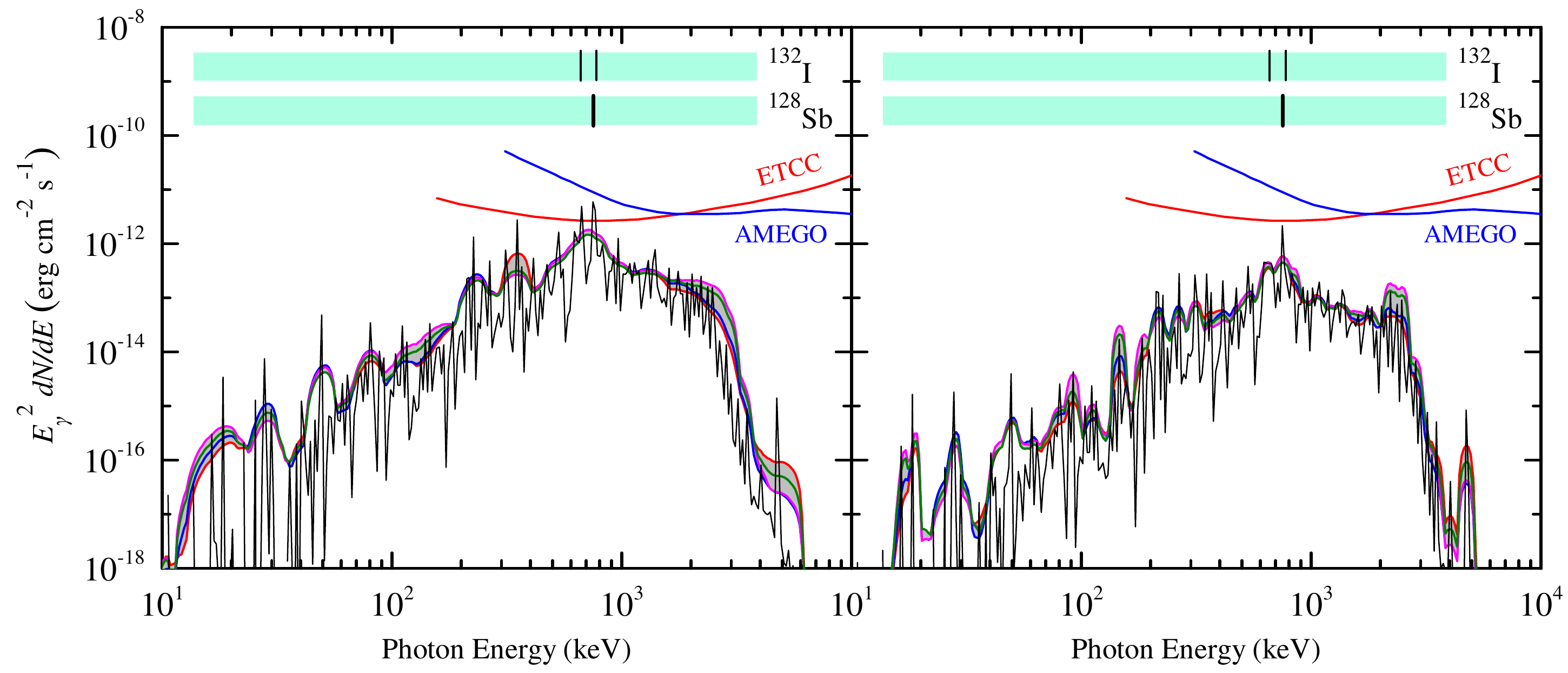}
\caption{The energy flux spectra associated with AT2017gfo derived from our model by adopting ejecta parameters as: {\em Left panel} --- $M_{\rm ej}=0.025M_{\odot}$ and $v_{\rm ej}=0.3c$ for the wind ejecta and $M_{\rm ej}=0.035M_{\odot}$ and $v_{\rm ej}=0.1c$ for the dynamical ejecta; {\em Right panel} --- $M_{\rm ej}=0.015M_{\odot}$ and $v_{\rm ej}=0.08c$ for the wind ejecta and $M_{\rm ej}=0.002M_{\odot}$ and $v_{\rm ej}=0.2c$ for the dynamical ejecta. The spectra are integrated over the first $10^6$~s. The grey band shows the range of resulting gamma-ray spectra from all our nuclear physics inputs. The sensitivity curves (exporsure time of $10^6$ seconds) of ETCC and AMEGO taken from \cite{Tanimori2015} and \cite{Moiseev2017} are plot for examining the delectability of the spectra with these proposed MeV gamma-ray missions. The dominant nuclide species ($^{132}$I and $^{128}$Sb) responsible for the line feature at around $800$~keV are also shown.}
\label{GW170817}
\end{figure}

In Figure~\ref{GW170817}, we show the derived GRT associated with AT2017gfo.
The energy flux spectrum is integrated over $10^6$~s starting from 1~hour after merger.
The derived spectra show several line features. The brightest one is around 800~keV, with a peak flux of $\sim10^{-12}$~erg~cm$^{-2}$~s$^{-1}$, which is generated by the radioactive decay of $^{128}$Sb and $^{132}$I.
The decay of $^{128}$Sb produces two bright gamma-ray lines with energy (intensity) of 743.3~keV ($100\%$) and 754.0~keV ($100\%$), while $^{132}$I produces gamma-ray lines at 667.7~keV ($98.7\%$) and 772.6~keV ($75.6\%$).

\section{Conclusion and Discussion}
\label{sec5}

We have studied the potential GRT powered by the radioactive decay of freshly-synthesized $r$-process elements in the ejecta of neutron star mergers with the two-component ejecta model, i.e. dynamical ejecta ($Y_{\rm e}\lesssim0.25$) and wind ejecta ($Y_{\rm e}\gtrsim0.25$).

Firstly, we analyze the composition and gamma-ray opacity of the ejecta.
The nuclear reaction network code SkyNet is adopted to calculate the evolution of the elemental abundances.
Based on the detailed nuclide composition of the ejected material, we analyzed the gamma-ray opacities of $r$-process elements by considering three main mechanisms of gamma-ray photons interacting with ejected material, i.e., photoelectric absorption, Compton scattering, and pair production, since these processes sensitively depends on the gamma-ray energy $E_{\gamma}$ and the atomic number $Z$ of the material. It is observed that the opacities are higher than that of the iron element in the sub-MeV energy range by a factor of $\sim4$ in the dynamical ejecta (composed of the heavy $r$-process elements) and $\sim 2$ in the wind ejecta (composed of the light $r$-process elements). This is because the probability of photoelectric absorption process is substantially enhanced by high-$Z$ elements.

Secondly, we investigate the potential GRTs based on the derived composition and gamma-ray opacity of the ejecta with the two-component model for ejecta mass and expansion velocity in the ranges of $M_{\rm ej}=0.001 \sim 0.05M_{\odot}$ and $v_{\rm ej}= 0.1\sim 0.4c$.
It is found that a more massive ejecta tends to power a later and brighter GRT, and a faster ejecta tends to power an earlier GRT. The dependence of the GRT peak time and peak luminosity generally follows the Equation~(\ref{tpk}) and (\ref{lpk}).
The value of $v_{\rm ej}$ can significantly affect the shape of the gamma-ray spectrum through the Doppler broadening effect.

Finally, as a case study, we apply our model to the possible GRT associated with AT2017gfo for evaluating its detection probability.
We adopt two different sets of ejecta parameters derived by several groups: (1) $M_{\rm ej}=0.025M_{\odot}$ and $v_{\rm ej}=0.3c$ for the wind ejecta and $M_{\rm ej}=0.035M_{\odot}$ and $v_{\rm ej}=0.1c$ for the dynamical ejecta; (2) $M_{\rm ej}=0.015M_{\odot}$ and $v_{\rm ej}=0.08c$ for the wind ejecta and $M_{\rm ej}=0.002M_{\odot}$ and $v_{\rm ej}=0.2c$ for the dynamical ejecta.
The derived spectra show several gamma-ray line features and the brightest one around 800~keV is generated by the radioactive decay of $^{132}$I and $^{128}$Sb.
This line feature can be used as a fingerprint to probe the production of the second $r$-process peak elements in future gamma-ray observation.

The flux threshold of the current MeV gamma-ray mission INTEGRAL is $5\times10^{-11}$~erg~cm$^{-2}$~s$^{-1}$ ($15$~keV$-8$~MeV, exposure time $10^6$~s, \citealp{Diehl2013}), being much higher than the peak gamma-ray energy fluxes derived from our analysis.
The proposed MeV gamma-ray instruments, such as ETCC ($0.15-20$~MeV, \citealp{Tanimori2015}) and AMEGO ($0.3$~MeV$-10$~GeV, \citealp{Moiseev2017}), have sensitivity of $2\times10^{-12}$~erg~cm$^{-2}$~s$^{-1}$ and $4\times10^{-12}$~erg~cm$^{-2}$~s$^{-1}$ for exposure time $10^6$~s, being closer to the peak gamma-ray flux of the GRT. For convincingly detecting such a GRT, the sensitivity of the instruments should be improved at least one order of magnitude.
To identify the gamma-ray line features, a detector having precent-level energy resolution would be most desired.
The proposed gamma-ray mission AMEGO can provide $<1\%$ energy resolution in the MeV band, but it is not sensitive enough for convincing detection of such a GRT at 40~Mpc.

We should emphasize that the detection of GRT event associated with AT2017gfo with ETCC and AMEGO like instruments is still over optimistic. The event rate of binary neutron star merger inferred from AT2017gfo is $R_{\rm BNS}=1540^{+3200}_{-1220}$~Gpc$^{-3}$~yr$^{-1}$ \citep{Abbott2017a}. The occurrence rates within a spherical volume of radius 40~Mpc is then $\sim0.4^{+0.9}_{-0.3}$~yr$^{-1}$. The extremely low occurrence rate and flux level make a great challenge to discover the GRT with ETCC and AMEGO like instruments. More sensitive instruments in the MeV gamma-ray band are required for increasing the detection probability of the GRT in a deeper universe.

\acknowledgments

We thank Li-Xin Li, Hou-Jun L$\ddot{\rm u}$, Da-Bin Lin, and Ning Wang for fruitful discussion and anonymous referee for helpful comments.
This work was supported by the National Natural Science Foundation of China (grant Nos.~12133003, 11773007, and U1731239) and the Guangxi Science Foundation (grant Nos.AD17129006, 2017AD22006, 2018GXNSFFA281010, and 2018GXNSFGA281007).

\newpage

\begin{thebibliography}{99}

\bibitem[Abbott et al.(2017a)]{Abbott2017a} Abbott, B.~P., Abbott, R., Abbott, T.~D., et al.\ 2017, \prl, 119, 161101. doi:10.1103/PhysRevLett.119.161101

\bibitem[Abbott et al.(2017b)]{Abbott2017b} Abbott, B.~P., Abbott, R., Abbott, T.~D., et al.\ 2017, \apjl, 848, L12. doi:10.3847/2041-8213/aa91c9

\bibitem[Arcavi et al.(2017)]{Arcavi2017} Arcavi, I., Hosseinzadeh, G., Howell, D.~A., et al.\ 2017, \nat, 551, 64. doi:10.1038/nature24291

\bibitem[Arnould et al.(2007)]{Arnould2007} Arnould, M., Goriely, S., \& Takahashi, K.\ 2007, \physrep, 450, 97. doi:10.1016/j.physrep.2007.06.002

\bibitem[Barnes \& Kasen(2013)]{Barnes2013} Barnes, J. \& Kasen, D.\ 2013, \apj, 775, 18. doi:10.1088/0004-637X/775/1/18

\bibitem[Barnes et al.(2016)]{Barnes2016} Barnes, J., Kasen, D., Wu, M.-R., et al.\ 2016, \apj, 829, 110. doi:10.3847/0004-637X/829/2/110

\bibitem[Bauswein et al.(2013)]{Bauswein2013} Bauswein, A., Goriely, S., \& Janka, H.-T.\ 2013, \apj, 773, 78. doi:10.1088/0004-637X/773/1/78

\bibitem[Beniamini et al.(2018)]{Beniamini2018} Beniamini, P., Dvorkin, I., \& Silk, J.\ 2018, \mnras, 478, 1994. doi:10.1093/mnras/sty1035

\bibitem[Burbidge et al.(1957)]{Burbidge1957} Burbidge, E.~M., Burbidge, G.~R., Fowler, W.~A., et al.\ 1957, Reviews of Modern Physics, 29, 547. doi:10.1103/RevModPhys.29.547

\bibitem[Chen et al.(2021)]{Chen2021} Chen, M.-H., Li, L.-X., Lin, D.-B., et al.\ 2021, \apj, 919, 59. doi:10.3847/1538-4357/ac1267

\bibitem[Chornock et al.(2017)]{Chornock2017} Chornock, R., Berger, E., Kasen, D., et al.\ 2017, \apjl, 848, L19. doi:10.3847/2041-8213/aa905c

\bibitem[C{\^o}t{\'e} et al.(2018)]{Cote2018} C{\^o}t{\'e}, B., Fryer, C.~L., Belczynski, K., et al.\ 2018, \apj, 855, 99. doi:10.3847/1538-4357/aaad67

\bibitem[Coulter et al.(2017)]{Coulter2017} Coulter, D.~A., Foley, R.~J., Kilpatrick, C.~D., et al.\ 2017, Science, 358, 1556. doi:10.1126/science.aap9811

\bibitem[Cowperthwaite et al.(2017)]{Cowperthwaite2017} Cowperthwaite, P.~S., Berger, E., Villar, V.~A., et al.\ 2017, \apjl, 848, L17. doi:10.3847/2041-8213/aa8fc7

\bibitem[Cyburt et al.(2010)]{Cyburt2010} Cyburt, R.~H., Amthor, A.~M., Ferguson, R., et al.\ 2010, \apjs, 189, 240. doi:10.1088/0067-0049/189/1/240

\bibitem[Diehl(2013)]{Diehl2013} Diehl, R.\ 2013, Reports on Progress in Physics, 76, 026301. doi:10.1088/0034-4885/76/2/026301

\bibitem[Dietrich et al.(2017)]{Dietrich2017} Dietrich, T., Ujevic, M., Tichy, W., et al.\ 2017, \prd, 95, 024029. doi:10.1103/PhysRevD.95.024029

\bibitem[Drout et al.(2017)]{Drout2017} Drout, M.~R., Piro, A.~L., Shappee, B.~J., et al.\ 2017, Science, 358, 1570. doi:10.1126/science.aaq0049

\bibitem[Duflo \& Zuker(1995)]{Duflo1995} Duflo, J. \& Zuker, A.~P.\ 1995, \prc, 52, R23. doi:10.1103/PhysRevC.52.R23

\bibitem[Evans et al.(2017)]{Evans2017} Evans, P.~A., Cenko, S.~B., Kennea, J.~A., et al.\ 2017, Science, 358, 1565. doi:10.1126/science.aap9580

\bibitem[Fujibayashi et al.(2018)]{Fujibayashi2018} Fujibayashi, S., Kiuchi, K., Nishimura, N., et al.\ 2018, \apj, 860, 64. doi:10.3847/1538-4357/aabafd

\bibitem[Goriely et al.(2008)]{Goriely2008} Goriely, S., Hilaire, S., \& Koning, A.~J.\ 2008, \aap, 487, 767. doi:10.1051/0004-6361:20078825

\bibitem[Hotokezaka et al.(2013)]{Hotokezaka2013} Hotokezaka, K., Kiuchi, K., Kyutoku, K., et al.\ 2013, \prd, 88, 044026. doi:10.1103/PhysRevD.88.044026

\bibitem[Hotokezaka et al.(2016)]{Hotokezaka2016} Hotokezaka, K., Wanajo, S., Tanaka, M., et al.\ 2016, \mnras, 459, 35. doi:10.1093/mnras/stw404

\bibitem[Hotokezaka et al.(2018)]{Hotokezaka2018} Hotokezaka, K., Beniamini, P., \& Piran, T.\ 2018, International Journal of Modern Physics D, 27, 1842005. doi:10.1142/S0218271818420051

\bibitem[Kasen et al.(2013)]{Kasen2013} Kasen, D., Badnell, N.~R., \& Barnes, J.\ 2013, \apj, 774, 25. doi:10.1088/0004-637X/774/1/25

\bibitem[Kasen et al.(2017)]{Kasen2017} Kasen, D., Metzger, B., Barnes, J., et al.\ 2017, \nat, 551, 80. doi:10.1038/nature24453

\bibitem[Kasliwal et al.(2017)]{Kasliwal2017} Kasliwal, M.~M., Nakar, E., Singer, L.~P., et al.\ 2017, Science, 358, 1559. doi:10.1126/science.aap9455

\bibitem[Kawaguchi et al.(2018)]{Kawaguchi2018} Kawaguchi, K., Shibata, M., \& Tanaka, M.\ 2018, \apjl, 865, L21. doi:10.3847/2041-8213/aade02

\bibitem[Kilpatrick et al.(2017)]{Kilpatrick2017} Kilpatrick, C.~D., Foley, R.~J., Kasen, D., et al.\ 2017, Science, 358, 1583. doi:10.1126/science.aaq0073

\bibitem[Kodama \& Takahashi(1975)]{Kodama1975} Kodama, T. \& Takahashi, K.\ 1975, \nphysa, 239, 489. doi:10.1016/0375-9474(75)90381-4

\bibitem[Kondev et al.(2021)]{Kondev2021} Kondev, F.~G., Wang, M., Huang, W.-J., et al.\ 2021, Chinese Physics C, 45, 030001. doi:10.1088/1674-1137/abddae

\bibitem[Korobkin et al.(2012)]{Korobkin2012} Korobkin, O., Rosswog, S., Arcones, A., et al.\ 2012, \mnras, 426, 1940. doi:10.1111/j.1365-2966.2012.21859.x

\bibitem[Korobkin et al.(2020)]{Korobkin2020} Korobkin, O., Hungerford, A.~M., Fryer, C.~L., et al.\ 2020, \apj, 889, 168. doi:10.3847/1538-4357/ab64d8

\bibitem[Lattimer \& Schramm(1974)]{Lattimer1974} Lattimer, J.~M. \& Schramm, D.~N.\ 1974, \apjl, 192, L145. doi:10.1086/181612

\bibitem[Li \& Paczy{\'n}ski(1998)]{Li1998} Li, L.-X. \& Paczy{\'n}ski, B.\ 1998, \apjl, 507, L59. doi:10.1086/311680

\bibitem[Li(2019)]{Li2019} Li, L.-X.\ 2019, \apj, 872, 19. doi:10.3847/1538-4357/aaf961

\bibitem[Lippuner \& Roberts(2015)]{Lippuner2015} Lippuner, J. \& Roberts, L.~F.\ 2015, \apj, 815, 82. doi:10.1088/0004-637X/815/2/82

\bibitem[Lippuner \& Roberts(2017)]{Lippuner2017} Lippuner, J. \& Roberts, L.~F.\ 2017, \apjs, 233, 18. doi:10.3847/1538-4365/aa94cb

\bibitem[Metzger et al.(2010)]{Metzger2010} Metzger, B.~D., Mart{\'\i}nez-Pinedo, G., Darbha, S., et al.\ 2010, \mnras, 406, 2650. doi:10.1111/j.1365-2966.2010.16864.x

\bibitem[Metzger(2019)]{Metzger2019} Metzger, B.~D.\ 2019, Living Reviews in Relativity, 23, 1. doi:10.1007/s41114-019-0024-0

\bibitem[Moiseev \& Amego Team(2017)]{Moiseev2017} Moiseev, A. \& Amego Team\ 2017, 35th International Cosmic Ray Conference (ICRC2017), 301, 798

\bibitem[M{\"o}ller et al.(2016)]{Moller2016} M{\"o}ller, P., Sierk, A.~J., Ichikawa, T., et al.\ 2016, Atomic Data and Nuclear Data Tables, 109, 1. doi:10.1016/j.adt.2015.10.002

\bibitem[Mumpower et al.(2018)]{Mumpower2018} Mumpower, M.~R., Kawano, T., Sprouse, T.~M., et al.\ 2018, \apj, 869, 14. doi:10.3847/1538-4357/aaeaca

\bibitem[Nedora et al.(2021)]{Nedora2021} Nedora, V., Bernuzzi, S., Radice, D., et al.\ 2021, \apj, 906, 98. doi:10.3847/1538-4357/abc9be

\bibitem[Nicholl et al.(2017)]{Nicholl2017} Nicholl, M., Berger, E., Kasen, D., et al.\ 2017, \apjl, 848, L18. doi:10.3847/2041-8213/aa9029

\bibitem[Pian et al.(2017)]{Pian2017} Pian, E., D'Avanzo, P., Benetti, S., et al.\ 2017, \nat, 551, 67. doi:10.1038/nature24298

\bibitem[Radice et al.(2018)]{Radice2018} Radice, D., Perego, A., Hotokezaka, K., et al.\ 2018, \apj, 869, 130. doi:10.3847/1538-4357/aaf054

\bibitem[Roederer(2017)]{Roederer2017} Roederer, I.~U.\ 2017, \apj, 835, 23. doi:10.3847/1538-4357/835/1/23

\bibitem[Rosswog(2005)]{Rosswog2005} Rosswog, S.\ 2005, \apj, 634, 1202. doi:10.1086/497062

\bibitem[Sekiguchi et al.(2016)]{Sekiguchi2016} Sekiguchi, Y., Kiuchi, K., Kyutoku, K., et al.\ 2016, \prd, 93, 124046. doi:10.1103/PhysRevD.93.124046

\bibitem[Shibata \& Hotokezaka(2019)]{Shibata2019} Shibata, M. \& Hotokezaka, K.\ 2019, Annual Review of Nuclear and Particle Science, 69, 41. doi:10.1146/annurev-nucl-101918-023625

\bibitem[Smartt et al.(2017)]{Smartt2017} Smartt, S.~J., Chen, T.-W., Jerkstrand, A., et al.\ 2017, \nat, 551, 75. doi:10.1038/nature24303

\bibitem[Sneden et al.(2008)]{Sneden2008} Sneden, C., Cowan, J.~J., \& Gallino, R.\ 2008, \araa, 46, 241. doi:10.1146/annurev.astro.46.060407.145207

\bibitem[Tanaka et al.(2017)]{Tanaka2017} Tanaka, M., Utsumi, Y., Mazzali, P.~A., et al.\ 2017, \pasj, 69, 102. doi:10.1093/pasj/psx121

\bibitem[Tanimori et al.(2015)]{Tanimori2015} Tanimori, T., Kubo, H., Takada, A., et al.\ 2015, \apj, 810, 28. doi:10.1088/0004-637X/810/1/28

\bibitem[Tanvir et al.(2017)]{Tanvir2017} Tanvir, N.~R., Levan, A.~J., Gonz{\'a}lez-Fern{\'a}ndez, C., et al.\ 2017, \apjl, 848, L27. doi:10.3847/2041-8213/aa90b6

\bibitem[Thielemann et al.(2017)]{Thielemann2017} Thielemann, F.-K., Eichler, M., Panov, I.~V., et al.\ 2017, Annual Review of Nuclear and Particle Science, 67, 253. doi:10.1146/annurev-nucl-101916-123246

\bibitem[Troja et al.(2017)]{Troja2017} Troja, E., Piro, L., van Eerten, H., et al.\ 2017, \nat, 551, 71. doi:10.1038/nature24290

\bibitem[Villar et al.(2017)]{Villar2017} Villar, V.~A., Guillochon, J., Berger, E., et al.\ 2017, \apjl, 851, L21. doi:10.3847/2041-8213/aa9c84

\bibitem[Wang et al.(2021)]{Wang2021} Wang, M., Huang, W.-J., Kondev F.~G., et al.\ 2021, Chinese Physics C, 45, 030003. doi:10.1088/1674-1137/abddaf

\bibitem[Watson et al.(2019)]{Watson2019} Watson, D., Hansen, C.~J., Selsing, J., et al.\ 2019, \nat, 574, 497. doi:10.1038/s41586-019-1676-3

\bibitem[Wollaeger et al.(2018)]{Wollaeger2018} Wollaeger, R.~T., Korobkin, O., Fontes, C.~J., et al.\ 2018, \mnras, 478, 3298. doi:10.1093/mnras/sty1018

\bibitem[Wu et al.(2019a)]{Wu2019a} Wu, M.-R., Banerjee, P., Metzger, B.~D., et al.\ 2019, \apj, 880, 23. doi:10.3847/1538-4357/ab2593

\bibitem[Wu et al.(2019b)]{Wu2019b} Wu, M.-R., Barnes, J., Mart{\'\i}nez-Pinedo, G., et al.\ 2019, \prl, 122, 062701. doi:10.1103/PhysRevLett.122.062701

\bibitem[Zhu et al.(2021)]{Zhu2021} Zhu, Y.~L., Lund, K.~A., Barnes, J., et al.\ 2021, \apj, 906, 94. doi:10.3847/1538-4357/abc69e

\end{thebibliography}
\end{document}